\documentclass[superscriptaddress,aps,prb,preprint]{revtex4-2}
\usepackage{bm}
\usepackage{dcolumn}
\usepackage{graphicx}
\usepackage{hyperref}
\usepackage{mathrsfs}
\usepackage{multirow}
\usepackage{rotating}
\usepackage{url}
\usepackage{color}
\usepackage{booktabs}
\usepackage{siunitx}

\newcommand{\Op}{{${\rm O_{3c}}$}}
\newcommand{\Onp}{{${\rm O_{4c}}$}}
\newcommand{\Vop}{{V$_{\rm O_{3c}}$}}
\newcommand{\Vonp}{{V$_{\rm O_{4c}}$}}

\newcommand{\etal}{{\em et al}.\ }

\newcommand{\note}[1]{{{#1}}} 
\newcommand{\notetwo}[1]{{{#1}}} 

\begin{document}


\title{A DFT+$U$+$V$ study of pristine and oxygen-deficient HfO$_2$ with self-consistent Hubbard parameters}

\author{Yudi Yang}
\affiliation{Zhejiang University, Hangzhou, Zhejiang 310058, China}
\affiliation{Department of Physics, School of Science and Research Center for Industries of the Future, Westlake University, Hangzhou, Zhejiang 310030, China}

\author{Wooil Yang}
\affiliation{Korea Institute for Advanced Study, Seoul 02455, Korea}

\author{Young-Woo Son}
\affiliation{Korea Institute for Advanced Study, Seoul 02455, Korea}

\author{Shi Liu}
\email{liushi@westlake.edu.cn}
\affiliation{Department of Physics, School of Science and Research Center for Industries of the Future, Westlake University, Hangzhou, Zhejiang 310030, China}
\affiliation{Institute of Natural Sciences, Westlake Institute for Advanced Study, Hangzhou, Zhejiang 310024, China}



\begin{abstract}

HfO$_2$-based ferroelectrics have emerged as promising materials for advanced nanoelectronics, with their robust polarization and silicon compatibility making them ideal for high-density, non-volatile memory applications. Oxygen vacancies, particularly in positively charged states, are suggested to profoundly impact the polymorphism kinetics and phase stability of hafnia, thereby affecting its ferroelectric behavior.
The electronic structures of pristine and oxygen-deficient hafnia polymorph have been extensively studied using density functional theory, primarily employing (semi-)local exchange-correlation functionals. However, these methods often underestimate band gaps and may not accurately capture the localized nature of $d$-electrons.
In this work, we investigate hafnia in various phases using DFT + $U$ + $V$, with onsite $U$ and intersite $V$ Hubbard parameters computed self-consistently via the pseudohybrid Hubbard density functional, ACBN0, and its extended version eACBN0. We find that the self-consistent DFT + $U$ method provides comparable accuracy to the computationally more expensive Heyd-Scuseria-Ernzerhof 
(HSE) hybrid density functional in predicting relative thermodynamic stability, band gaps, and density of states. Furthermore, it is a cost-effective approach for estimating the formation energies of oxygen vacancies. Additionally, we demonstrate that environmentally dependent Hubbard parameters serve as useful indicators for analyzing bond strengths and electronic structures in real space. 
\end{abstract}

\maketitle

\newpage
\section{Introduction}
The discovery of ferroelectricity in HfO$_2$-based thin films in 2011~\cite{Boscke11p102903, Luo20p1391} has opened up an avenue to realize nanoelectronics such as non-volatile memory and memristor, as this fluorite-structured material demonstrates significant advantages over conventional perovskite ferroelectrics. Specifically, HfO$_2$-based ferroelectrics exhibit robust switchable polarization in thin films with a thickness down to 1~nm~\cite{Schroeder22p653, Park15p1811}, which is highly desirable for the development of high-density memory devices. In addition, the excellent compatibility of HfO$_2$ with complementary metal oxide semiconductor (CMOS) technology offers the potential for seamless integration of ferroelectric functionalities with existing silicon-based electronic devices~\cite{Stefon09p402}. These unique properties of ferroelectric HfO$_2$ have thus sparked intense research interest, motivating the development of new techniques for its synthesis, characterization, and integration into electronic devices. 

\begin{figure}[h]
    \centering
    \includegraphics[width=1\textwidth]{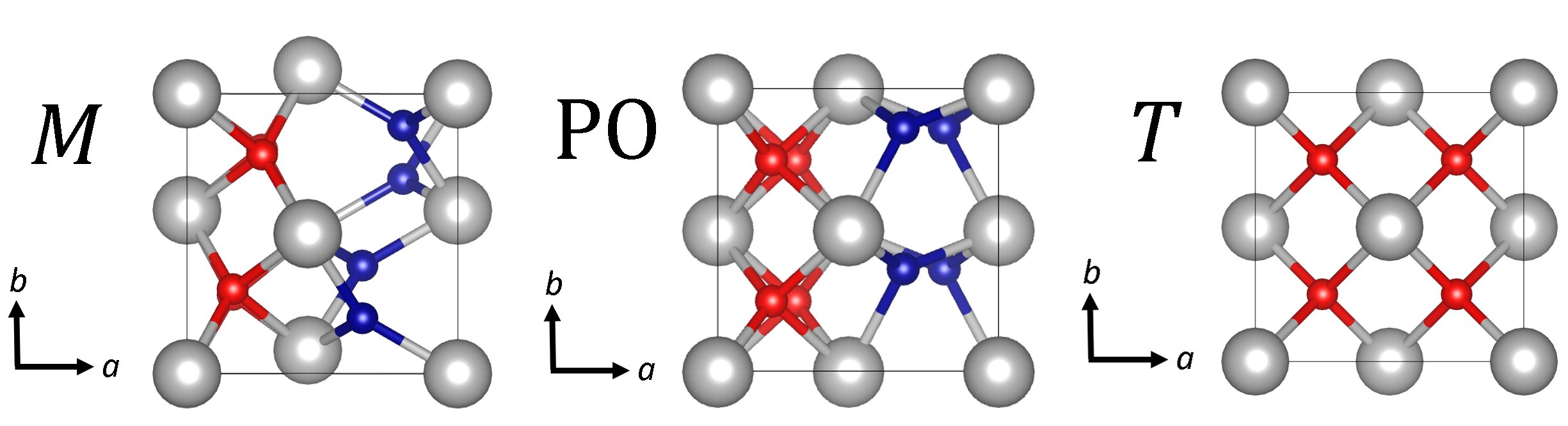}
    \caption{Schematics of $P2_1/c$ ($M$), $Pca2_1$ (PO) and $P4_2/nmc$ ($T$) phases of HfO$_2$. Three-fold and four-fold coordinated oxygen atoms are colored in blue and red, respectively.}
    \label{structure}
\end{figure}

The polar orthorhombic (PO) $Pca2_1$ phase of HfO$_2$ is widely recognized as the origin of ferroelectricity in hafnia thin films~\cite{Zhang21p135, Huan14p064111, Sang15p162905}, despite being energetically less favorable than the nonpolar monoclinic ($M$) $P2_1/c$ phase (see Fig.~\ref{structure})~\cite{Materlik15p134109}. 
Among various extrinsic factors that can affect the relative stability between the PO and $M$ phase in thin films of hafnia~\cite{Park15p1811}, oxygen vacancies (V$_{\rm{O}}$) have been suggested to play a crucial role in stabilizing the ferroelectric phase, either thermodynamically or kinetically, in a number of experimental and theoretical studies~\cite{Chouprik21p11635, Guha07p196101, Cho08p233118, Zhou19p143}. Particularly, first-principles density functional theory (DFT) calculations have revealed the drastically different impacts of charge-neutral and charged oxygen vacancies on the relative stability of HfO$_2$ polymorphs, which are intimately related to the energy of the defect level~\cite{He21pL180102} and the degree of electronic screening around the vacancy~\cite{Ma23p096801}. However, previous DFT investigations on HfO$_2$ have mainly relied on (semi-)local exchange-correlation (XC) functionals such as the local density approximation (LDA)~\cite{Perdew92p13244} and generalized gradient approximation (GGA)~\cite{Perdew96p3865}. These methods are known to have limitations in predicting the electronic properties of materials with localized $d$ or $f$ states, such as those containing transition-metal elements, due to the remnant self-interaction error (SIE)~\cite{Perdew81p5048}. This could pose a potential issue for accurate descriptions of the electronic structures of HfO$_2$, which involves localized $d$ states. To achieve a correct understanding of how V$_{\rm{O}}$ of varying charge states influence properties of HfO$_2$-based ferroelectrics, it is worth revisiting the electronic and structural properties of HfO$_2$, both with and without V$_{\rm{O}}$, using more accurate first-principles methods.



Hybrid XC functionals, while mitigating SIE by mixing a fraction ($\alpha$) of Hartree-Fock (HF) exact exchange with the semilocal DFT exchange, nevertheless result in high computational costs for point-defect calculations requiring large supercells. The PBE0 hybrid functional combines 75\% of Perdew-Burke-Ernzerhof (PBE) exchange with a fixed 25\% HF exchange ($\alpha=0.25$)~\cite{Adamo99p6158, Perdew96p9982}. Heyd, Scuseria, and Ernzerhof developed the screened hybrid functional HSE06~\cite{Heyd03p8207, Paier06p154709, Marsman08p064201}, wherein the short-range exchange includes a portion of HF exchange ($\alpha=0.25$), while the long-range exchange is treated with PBE exchange only~\cite{Yang23p108}. In some studies using HSE06, the value of $\alpha$ is treated as a material-specific parameter that characterizes the strength of dielectric screening and is empirically tuned to align with experimental observations~\cite{Franchini14p253202}. This approach, though improving the agreement between experiment and theory, inevitably sacrifices the {\em ab initio} nature of DFT calculations.
It is worth noting that the HF mixing parameter $\alpha$ can also be determined self-consistently~\cite{Skone14p195112}. The self-consistent cycle begins with an initial guess for $\alpha$ and iteratively computes the static dielectric constant using the hybrid exchange-correlation potential defined by the $\alpha$, updating $\alpha$ until convergence for the static dielectric constant is achieved. 
By computing $\alpha$ directly from the electronic structure of the system, the first-principles nature of the calculations is preserved, eliminating the need for empirically tuning the parameter and enhancing the predictive power of the approach. 

The $GW$ approximation is a many-body perturbation theory method used to calculate the electronic structure of materials, particularly quasiparticle energies such as band gaps~\cite{Hedin69book, Aryasetiawan98p237, Hedin65p796, Hybertsen86p5390}. It improves upon standard DFT by explicitly incorporating electron-electron interactions through the self-energy, which is approximated as the product of the Green's function ($G$) and the screened Coulomb interaction ($W$), typically computed within the random phase approximation (RPA).
The simplest form of $GW$, known as $G_0W_0$, is a single-shot calculation where quasiparticle energies are obtained from a single $GW$ iteration. In this approach, all off-diagonal matrix elements of the self-energy are neglected, and a Taylor expansion of the self-energy is performed around the DFT eigenvalues. A more refined approach, called $GW_0$, introduces partial self-consistency by iteratively updating the Green's function ($G$) while keeping the screened Coulomb interaction ($W_0$) fixed~\cite{Shishkin07p235102}. This improves the accuracy of the quasiparticle energies while maintaining computational efficiency.

The DFT + $U$ approach and the extended version, DFT + $U$ + $V$, both based on the mean-field Hubbard model, offer computationally efficient ways to correct SIE, yielding improved descriptions of electronic structures for transition-metal systems compared to LDA and GGA functionals~\cite{Anisimov91p943, Liechtenstein95pR5467, Anisimov97p767, Dudarev98p1505, Kulik06p103001}. In these approaches, a Hubbard functional is incorporated into standard DFT, accounting for Coulomb repulsion between strongly localized electrons of the same atom, as well as Coulomb interactions between electrons on neighboring sites, which are determined by the ``onsite" Hubbard $U$ parameter and ``intersite" Hubbard $V$ parameter, respectively~\cite{LeiriaCampoJr10p055602}. 
There are several main methods reported in the literature for determining the Hubbard $U$ and $V$ parameters. First, a common practice is to assume these Hubbard parameters are element-specific and to tune them to match experimental results or results obtained with higher-level theories such as HSE06 and $GW$ approximations. This empirical approach, similar to tuning $\alpha$ in hybrid functionals, compromises the first-principles aspect of DFT calculations. Moreover, the values of Hubbard parameters vary significantly when targeting different experimental properties, leading to discrepancies in predicting properties such as defect energetics~\cite{Wang06p195107, Lutfalla11p2218, Getsoian13p25562, Capdevila-Cortada16p8370}. 
Therefore, it is preferable to determine the Hubbard parameters in a systematic and self-consistent way. Second, the linear response constrained density functional theory (LR-cDFT) interprets $U$ and $V$ as corrections required to restore the piece-wise linear relationship between the total energy and orbital occupation, where their values are extracted from the response matrix of orbital occupation to a perturbing potential. This method can be computationally demanding due to the need for supercell calculations and might be numerically unstable, as it could yield unphysically large $U$ values for fully occupied orbitals~\cite{Lee12p781}. Third, Timrov \etal~reformulated LR-cDFT with density functional perturbation theory (DFPT), which recasts perturbations in supercells as a sum of wavevector-specific perturbations in a primitive cell in reciprocal space, enabling efficient calculation of site-dependent Hubbard parameters without needing supercell calculations. 
\note{Fourth, the constrained random phase approximation (cRPA) method can be used to calculate the Hubbard $U$ by determining the screened Coulomb interaction within a specified subspace~\cite{Pavarini11book}, typically utilizing the polarization function of the system. However, this method can encounter challenges when applied to 3$d$ transition metals due to the significant entanglement of the 3$d$ bands with other bands, such as the 4$s$ and 4$p$ bands. This entanglement complicates the isolation of the 3$d$ subspace, potentially leading to inaccuracies in the calculated $U$ values and instability in the results~\cite{Pavarini11book}.}
Lastly, the pseudohybrid Hubbard density functional, namely Agapito-Curtarolo-Buongiorno Nardelli (ACBN0)~\cite{Agapito15p011006} and its extended version, eACBN0~\cite{LeiriaCampoJr10p055602, Lee20p043410, Yang21p104313, Yang24p155133}, introduces a self-consistent method for calculating $U$ and $V$ values. The HF energy associated with a chosen Hubbard manifold is expressed in terms of renormalized density matrices and occupations, leading to density functionals of $U$ and $V$ that are updated during an electronic self-consistent cycle. 

The ACBN0 and eACBN0 approaches have demonstrated high accuracy compared to the PBE functional and represent a more computationally efficient alternative to HSE06 and $GW_0$~\cite{Hedin65p796}. Inherent to the Hubbard model, the $U$ and $V$ parameters should depend on the local atomic environment and thus be Hubbard-site specific rather than element-specific. This characteristic is correctly captured within both ACBN0 and eACBN0. 
It should be noted that since ACBN0 and eACBN0 are relatively new density functionals, few studies have investigated their performance in predicting structural properties and defect formation energies~\cite{Ricca20p023313,Timrov21p045141}, which fundamentally require calculations of relative energy differences. The impact of local environment-sensitive Hubbard parameters on predicting relative energy differences between different configurations deserves further investigation.




This study, taking HfO$_2$ as a paradigmatic case, compares the results from the ACBN0 and eACBN0 functionals with those obtained from HSE06 and $GW$.  In ACBN0, the Hubbard model incorporates onsite interactions of Hf-5$d$ and O-2$p$ states, while eACBN0 includes additional intersite interactions between these states. We systematically analyze the electronic structures of pristine and oxygen-deficient HfO$_2$ obtained with self-consistent DFT + $U$ + $V$, by comparing the results with those obtained with HSE06 and $GW_0$. 
\note{It is noted that we adopt the standard ``off-the-shelf" value of 0.25 for $\alpha$ without any tuning, as our goal is to minimize ``human involvement" in this benchmark study. We select the HSE06 hybrid functional primarily for its practical convenience and widespread recognition as one of the most accurate hybrid functionals for solid-state materials~\cite{Deak10p153203, Yang23p108, Seidl21p034602}.
In the specific case of HfO$_2$, the HSE06 functional has demonstrated excellent accuracy in predicting bandgap values. (see discussions below).}
Environmental-dependent Hubbard parameters are employed to assess the impact of oxygen vacancies on the electronic structures in real space. 
We calculate the oxygen vacancy formation energies for various phases of HfO$_2$ and compare the values predicted by different functionals. It is
demonstrated that ACBN0 offers comparable accuracy to the computationally more expensive \note{HSE06 in predicting relative thermodynamic stability, band gaps, and density of states. Additionally, ACBN0 achieves a similar level of accuracy as $GW_0$ in the calculation of the density of states.}
This comparative analysis provides useful insights into the ACBN0 and eACBN0 functionals and their applicability in modeling functional oxides.


\section{Computational Methods}
We consider three experimentally observed phases of HfO$_2$: $M$, PO, and the high-temperature tetragonal ($T$) phase (space group $P4_2/nmc$). All DFT calculations are performed with QUANTUM ESPRESSO (QE)~\cite{Giannozzi09p395502, Giannozzi17p465901} using Optimized Norm-Conserving Vanderbilt pseudopotentials taken from the PseudoDojo library~\cite{Van18p39} 
The structural parameters of 12-atom unit cells are optimized with the PBE functional using a plane wave cutoff energy of 60 Ry, a 4$\times$4$\times$4 \textit{k}-point mesh for Brillouin zone sampling, a Gaussian smearing of 0.01 Ry, an energy convergence threshold of 10$^{-4}$ Ry, and a force convergence threshold of 10$^{-3}$ Ry/Bohr. Both lattice constants and ionic positions are fully relaxed. For density of states (DOS) calculations, we employ the tetrahedra smearing and an 8$\times$8$\times$8 $k$-point mesh.

Based on PBE-optimized structures, we compute $U$ and $V$ parameters self-consistently with ACBN0 and eACBN0 using an in-house version of QE. The on-site $U$ corrections are applied to Hf-5$d$ states and O-2$p$ states. The intersite Hubbard interactions between nearest-neighboring Hf-5$d$ and O-2$p$ are included in DFT + $U$ + $V$. The Hubbard $U$ and $V$ parameters are considered converged when the change in the Hubbard energy is within 10$^{-5}$ Ry.

The defective hafnia systems with oxygen vacancies are modeled using 2$\times$2$\times$2 supercells with one oxygen atom removed. Different charge states ($q$) of V$_{\rm O}$, denoted as V$^{\times}_{\rm{O}}$ for $q=0$, V$^{\bullet}_{\rm{O}}$ for $q=+1$, and V$^{\bullet\bullet}_{\rm{O}}$ for $q=+2$ in Kr\"{o}ger-Vink notation~\cite{Krger56p307}, are created by adjusting the number of electrons of the system. The background-charge method is adopted when modeling a charged vacancy.
The ionic positions are optimized with PBE, while the lattice constants are fixed at the ground-state values of unit cells, to stimulate the dilute defect limit. The formation energy of an oxygen vacancy in charge state $q$ is calculated following the methodology outlined in Ref.~\cite{Naik18p114}, employing the Freysoldt–Neugebauer–Van de Walle (FNV) correction scheme~\cite{Freysoldt09p016402, Freysoldt14p253}:
\begin{equation}
    E^{\rm{f}}_{q}(\epsilon_{\rm{F}}) = \{E^{\rm{tot}}_{q} + E_{q}^{\rm{corr}}\} - E_{\rm{pristine}} + q\{\epsilon^{\rm{pristine}}_{\rm{vbm}} + \epsilon_{\rm{F}} - \Delta V_{0/\rm{p}}\} - n_{\rm{O}}\mu_{\rm{O}},
    \label{Vo formation energy charge}
\end{equation}
where $E^{\rm{tot}}_{q}$ is the total energy of the supercell containing a charged vacancy and $E_{\rm{pristine}}$ is the total energy of a pristine supercell of the same size; $E_{q}^{\rm{corr}}$ is the finite-size electrostatic correction, which amends 
the spurious interaction between the defect charge and its periodic images arising from the usage of periodic boundary conditions; $\epsilon_{\rm{F}}$ is the Fermi level with respect to the valence band maximum (VBM) in the pristine supercell, $\epsilon^{\rm{pristine}}_{\rm{vbm}}$; $\Delta V_{0/\rm{p}}$ is the potential alignment term obtained by comparing the electrostatic potentials far from a defect of $q=0$ ($V_{0}|_{\rm{far}}$) and that in a pristine supercell ($V_{\rm{p}}$), that is, $\Delta V_{0/\rm{p}} = V_{0}|_{\rm{far}} - V_{\rm{p}}$; For the chemical potential of the oxygen atom, $\mu_{\rm{O}}$, it is set as the half of the energy of an oxygen molecule for oxygen-rich conditions. 
For a neural defect of $q=0$, the absence of long-range electrostatic defect-defect interactions simplifies the determination of the formation energy:
\begin{equation}
    E^{\rm{f}}_{\rm{0}}(\epsilon_{\rm{F}}) = E^{\rm{tot}}_{\rm{0}} - E_{\rm{pristine}} - n_{\rm{O}}\mu_{\rm{O}}.
    \label{Vo formation energy neutral}
\end{equation}

\section{Results and Discussion}
\subsection{Energetics of HfO$_2$ polymorphs}
The unit cells of hafnia polymorphs (in Fig~\ref{structure}) are first optimized using the PBE functional, followed by single-point energy calculations using ACBN0, eACBN0, and HSE06. 
\note{We also performed geometry optimizations using ACBN0 and eACBN0, following the workflow outlined in Ref.~\cite{Yang22p195159}. The structural parameters obtained with ACBN0 and eACBN0 show good agreement with PBE values for the $M$ and PO phases. While ACBN0 yields structural parameters for the $T$ phase that are nearly identical to the PBE values, we encountered difficulties in optimizing the $T$ phase with eACBN0, as the structure tends to relax into other lower-energy phases. This behavior is expected, as the $T$ phase is a high-energy metastable phase.}
The lattice constants used in this paper and self-consistent Hubbard $U$ computed with ACBN0 and eACBN0 are reported in Table~\ref{Acbn0U}. Unlike the higher-symmetry $T$ phase, where all oxygen atoms are symmetrically equivalent and bonded to four hafnium atoms, the structures of the $M$ and PO phases feature alternating fourfold-coordinated (O$_{\rm 4c}$) and threefold-coordinated (O$_{\rm 3c}$) oxygen atoms. The variation in the local environments of oxygen atoms is reflected in the self-consistent $U$ values. Both ACBN0 and eACBN0 predict slightly higher $U$ values for the $2p$ states of O$_{\rm 3c}$ atoms, $\approx$8.8 eV for ACBN0 and $\approx$8.5 eV for eACBN0, compared to the ACBN0 value of $\approx$8.2 eV and eACBN0 value of $\approx$8.0 eV for the $2p$ states of O$_{\rm 4c}$ atoms.
The $U$ values for Hf-5$d$ states are small ($<$0.3~eV), indicating these localized states are already adequately described by PBE.

\begin{figure*}[h]
    \centering
    \includegraphics[]{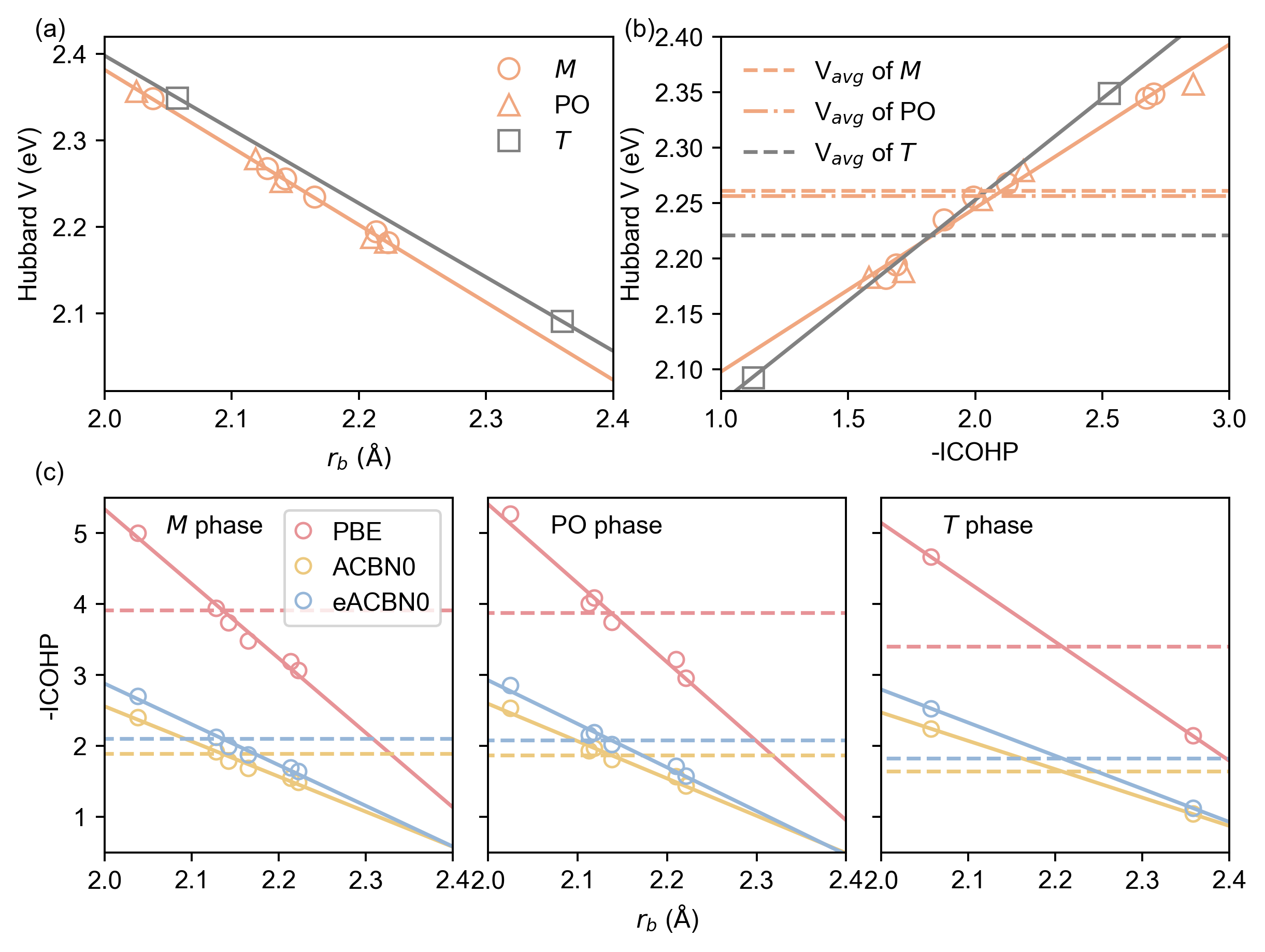}
    \caption{(a) Relationship between the Hubbard $V$ and the Hf-O bond length ($r_b$). (b) Relationship between the Hubbard $V$ and bond strength, represented by $-$ICOHP. (c) Correlation between $r_b$ and $-$ICOHP calculated using different functionals. Dashed lines indicate the average magnitude of $-$ICOHPs.}
    \label{VBond}
\end{figure*}

The diverse Hf-O bonds present in hafnia polymorphs allow for quantitative analysis of the relationship between Hf-O bond length ($r_b$) and the self-consistent intersite Hubbard $V$ values computed using eACBN0. As depicted in Fig.~\ref{VBond}(a), the magnitude of $V$ is inversely proportional to $r_b$ across all phases. Notably, the $M$ and PO phases demonstrate nearly identical $V$-$r_b$ relationships, with $V$ values ranging from 2.20 to 2.35 eV. 
In contrast, the $V$ values for the two Hf-O bonds in the $T$ phase show a more substantial variation, differing by $\approx$0.25 eV. The linear regression of $V$ versus $r_b$ in the $T$ phase indicates a slightly higher $V$ value for the same $r_b$ when compared to those in the $M$ and PO phases. 

To elucidate the impact of Hubbard parameters on bonding strength, we compute the ICOHPs for each Hf-O bond in various polymorphs of hafnia. 
\note{The Crystal Orbital Hamilton Population (COHP) method~\cite{Dronskowski93p8617} reformulates the basis of a band structure energy partitioning scheme into a sum of orbital pair contributions, providing a detailed analysis of chemical bonding by decomposing the electronic structure into atom- and bond-specific components. A COHP versus energy diagram reveals bonding, nonbonding, and antibonding energy regions across a specified energy range. The integrated COHP (ICOHP) is the energy integral of the COHP over a chosen energy range, and it serves as a quantitative measure of the bonding strength between atoms or within bonds.}
The correlation between the Hubbard parameter $V$ and the negative value of ICOHP ($-$ICOHP), as depicted in Fig.~\ref{VBond}(b), shows a positive relationship, suggesting that the magnitude of $V$ could serve as a useful indicator of bond strength.
We observe that the $V$-$r_b$ relationship in the $T$ phase exhibits a distinct slope compared to those in the $M$ and PO phases. This suggests that while the chemical bonding in the $M$ and PO phases is similar, it differs from that in the $T$ phase.

A comparative analysis of bond strength across different functionals, as presented in Fig.~\ref{VBond}(c), reveals that the PBE functional predicts the strongest bonds among the three considered functionals. The introduction of the Hubbard $U$ parameter in ACBN0 tends to localize electrons, which generally weakens the interactions between neighboring atoms.  
Conversely, the Hubbard $V$ parameter is associated with interatomic interactions that enhance electronic coupling between neighboring atoms, thereby strengthening the bonds. This explains why eACBN0 predicts stronger Hf-O bonds than ACBN0.  Furthermore, a consistent trend in bond strength emerges across all functionals: the $M$ phase exhibits the highest average bond strength, followed by the PO phase with moderate bond strength, while the $T$ phase is characterized by the lowest bond strength. This hierarchy in bond strength aligns with the thermodynamic stability order of these phases, which will be discussed in more detail.

\begin{figure}[h]
    \centering
    \includegraphics[]{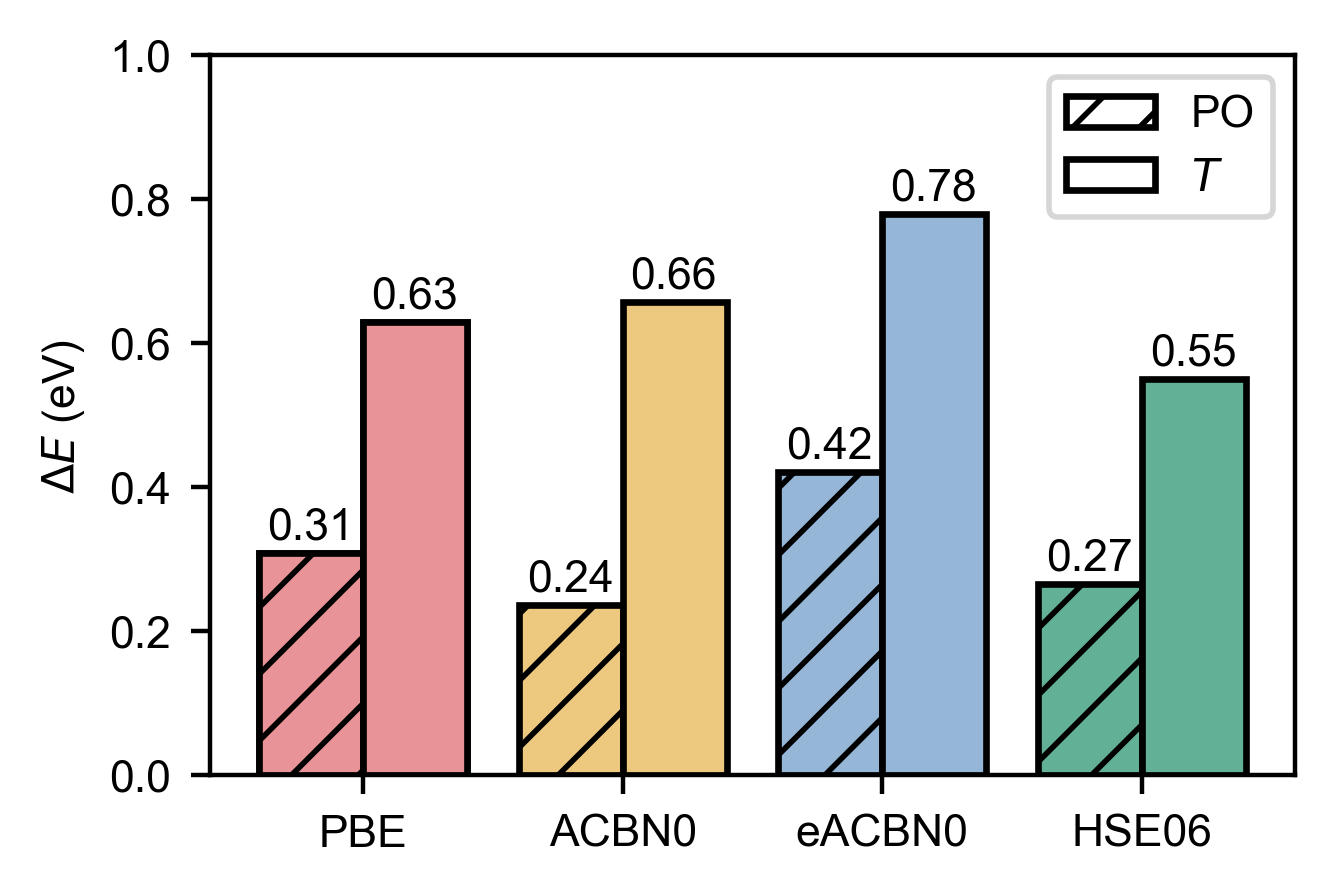}
    \caption{Relative energies of PO and $T$ phases predicted with different functionals. The energy of the $M$ phase is set to zero as a reference.}
    \label{energy_diff}
\end{figure}

Figure~\ref{energy_diff} plots the relative energies of PO and $T$ phases in reference to the $M$ phase, as computed with PBE, ACBN0, eACBN0, and HSE06 functionals. All functionals predict the same energetic ordering, with the energies of $M$, PO, and $T$ phases increasing sequentially. The energy differences predicted by PBE, ACBN0, and HSE06 are comparable, whereas eACBN0 yields significantly larger values. To elucidate the predominant factors influencing these energy differences, we analyze the total energy by breaking it down into individual contributions: one-electron, Hartree, exchange-correlation, Ewald, and Hubbard energies. 
It is evident from Table~\ref{energy_separate} that the larger energy difference predicted by eACBN0 mainly stems from a more significant variation in the Hubbard energy between the hafnia polymorphs.

\subsection{Electronic structures of HfO$_2$ polymorphs}

Before the discovery of ferroelectricity in hafnia-based thin films, hafnia had already attracted significant interest as promising high-$\kappa $ materials.
The electronic properties of HfO$_2$ have been extensively studied using various experimental techniques, including ultraviolet plus inverse photoemission spectroscopy~\cite{Sayan04p7485, Bersch08p085114}, spectroscopic ellipsometry~\cite{Cho02p1249, Modreanu03p1236}, electron-energy-loss spectroscopy~\cite{Yu02p376, Puthenkovilakam04p2701}, and x-ray absorption spectroscopy~\cite{Lucovsky04p288}. Reported band gaps ($E_g$) range from 5 to 6 eV, due to variations in the samples, which may be crystalline or amorphous. Even within crystalline films, different phases may be present, contributing to the variability of the band gaps. We compute the band gaps of hafnia polymorphs using four different density functionals and compare them with the results obtained from the high-level method represented by $GW_0$~\cite{Jiang10p085119}.

\begin{figure*}[h]
    \includegraphics[]{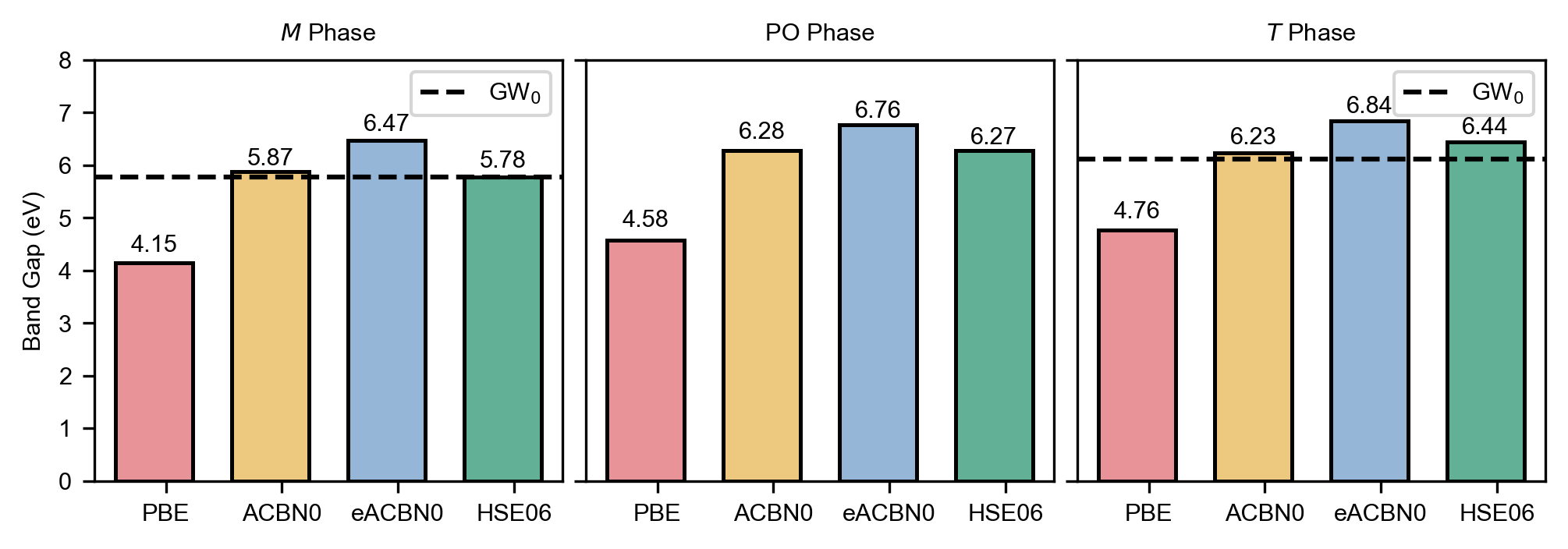}
    \caption{Band gap values of different phases of HfO$_2$ calculated using different functionals,  compared to $GW_0$ results reported in ref.~\cite{Jiang10p085119}. }
    \label{bandgap}
\end{figure*}

As shown in Fig.~\ref{bandgap}, for the $M$ phase, PBE predicts a band gap of 4.15 eV, while ACBN0 and HSE06 give higher values of 5.87 eV and 5.78 eV, respectively, aligning well with the experimental result of 5.68~eV~\cite{Balog77p247} and the $GW_0$ estimate of 5.78 eV~\cite{Jiang10p085119}. In contrast, eACBN0 with self-consistent $U$ and $V$ parameters yields the largest gap of 6.47~eV. A similar trend is observed for the PO and $T$ phases, where ACBN0 and HSE06 predict band gaps comparable to the $GW_0$ results, whereas PBE estimates are lower and eACBN0 values are higher. Overall, ACBN0, with self-consistently determined Hubbard $U$ parameters, achieves a level of accuracy comparable to HSE06 and $GW_0$ but with greater computational efficiency. Additionally, as detailed below, the large band gap predicted by eACBN0 has a surprisingly strong impact on the formation energy of charge-neutral oxygen vacancies.

\begin{figure*}[h]
    \includegraphics[]{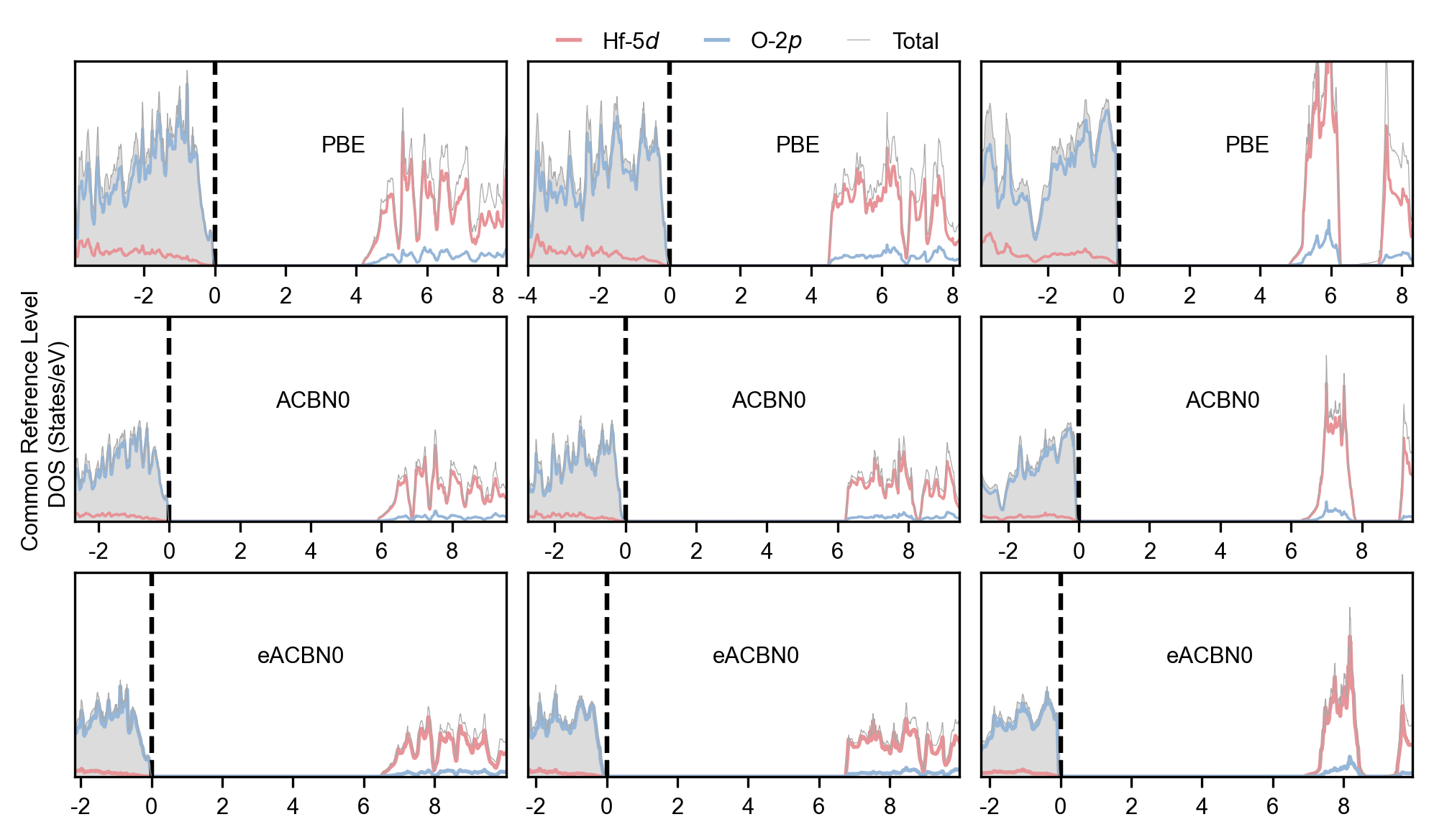}
    \caption{Density of states (DOS) for different phases ($M$, PO, and $T$, from left to right) calculated using various functionals. The DOS plots for each phase are aligned at the core energies.}
    \label{dos_allcal}
\end{figure*}

The effects of Hubbard parameters on the electronic structures are analyzed by comparing the density of states (DOS). In all three polymorphs, the primary contribution to the valence band maximum (VBM) comes from the O-2$p$ state, while the conduction band minimum (CBM) mainly involves the Hf-5$d$ state. As illustrated in Fig.~\ref{dos_allcal}, we align the DOS plots computed with different functionals based on their core electron energies (assuming core electrons are less affected by \note{Hubbard parameters}). This alignment allows for a direct comparison of absolute energy levels. 
Compared to PBE, the introduction of Hubbard $U$ corrections of $\approx$8.5~eV to O-2$p$ states in ACBN0 lowers the VBM. In eACBN0, the incorporation of Hubbard $V$ interactions between Hf-5$d$ and O-2$p$ states further decreases the VBM energy. In contrast, the CBM shows less sensitivity to Hubbard corrections, likely due to the much smaller $U$ corrections of $\approx$0.2~eV applied to Hf-$5d$ states. \note{This trend is consistent with the
consecutively increasing band gap values predicted by PBE, ACBN0, and
eACBN0.}

\begin{figure}[h]
    \includegraphics[]{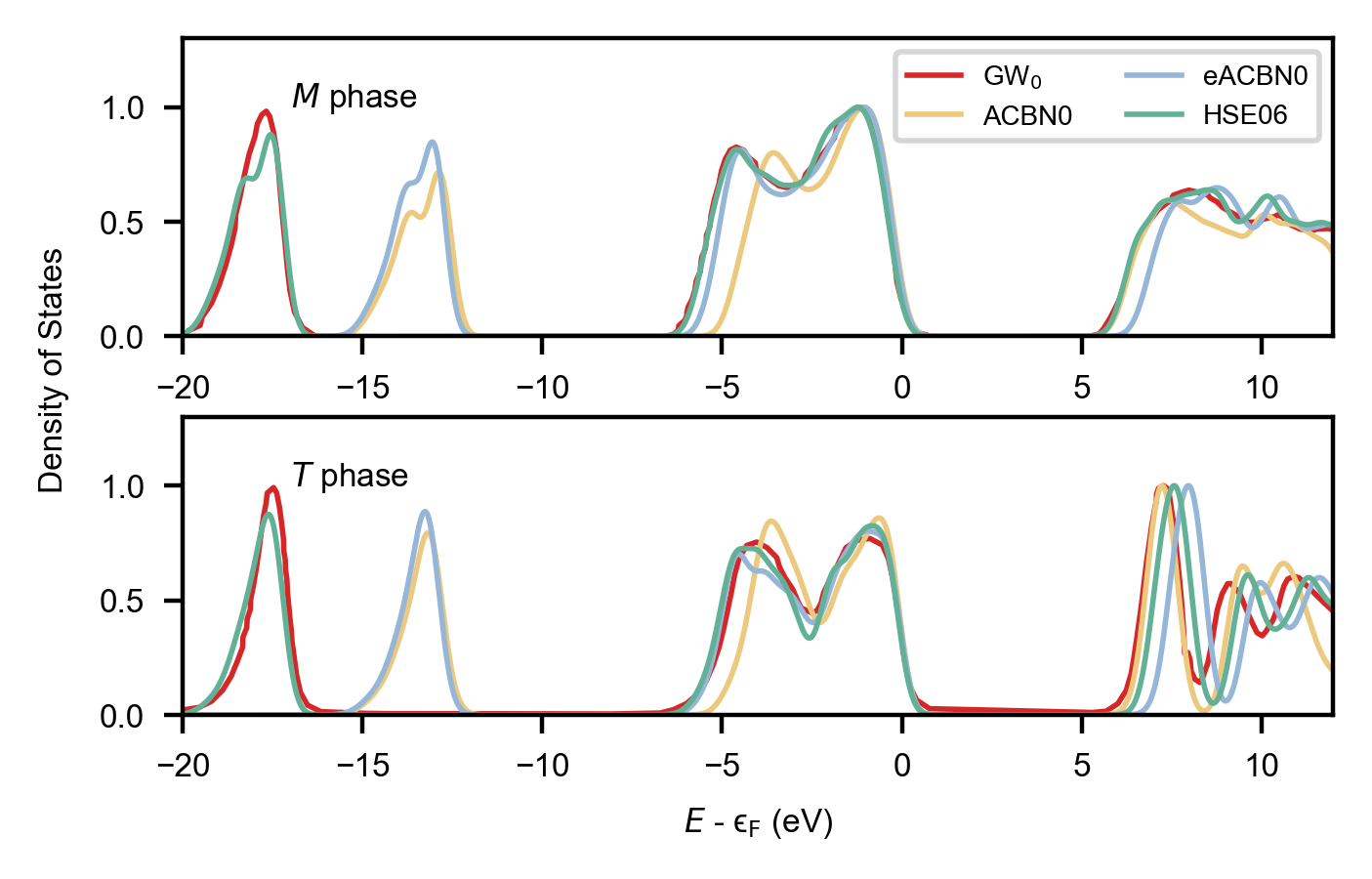}
    \caption{\notetwo{Comparison of DOS plots computed using ACBN0, eACBN0, HSE06, and $GW_0$~\cite{Jiang10p085119} for $M$ and $T$ phase, with the VBM aligned. Notice that the gap predicted by ACBN0, HSE06, and $GW_0$ are similar, but eACBN0 predicted a larger gap than the others.} }
    \label{gwdos}
\end{figure}

\begin{figure*}[h]
    \centering
    \includegraphics[]{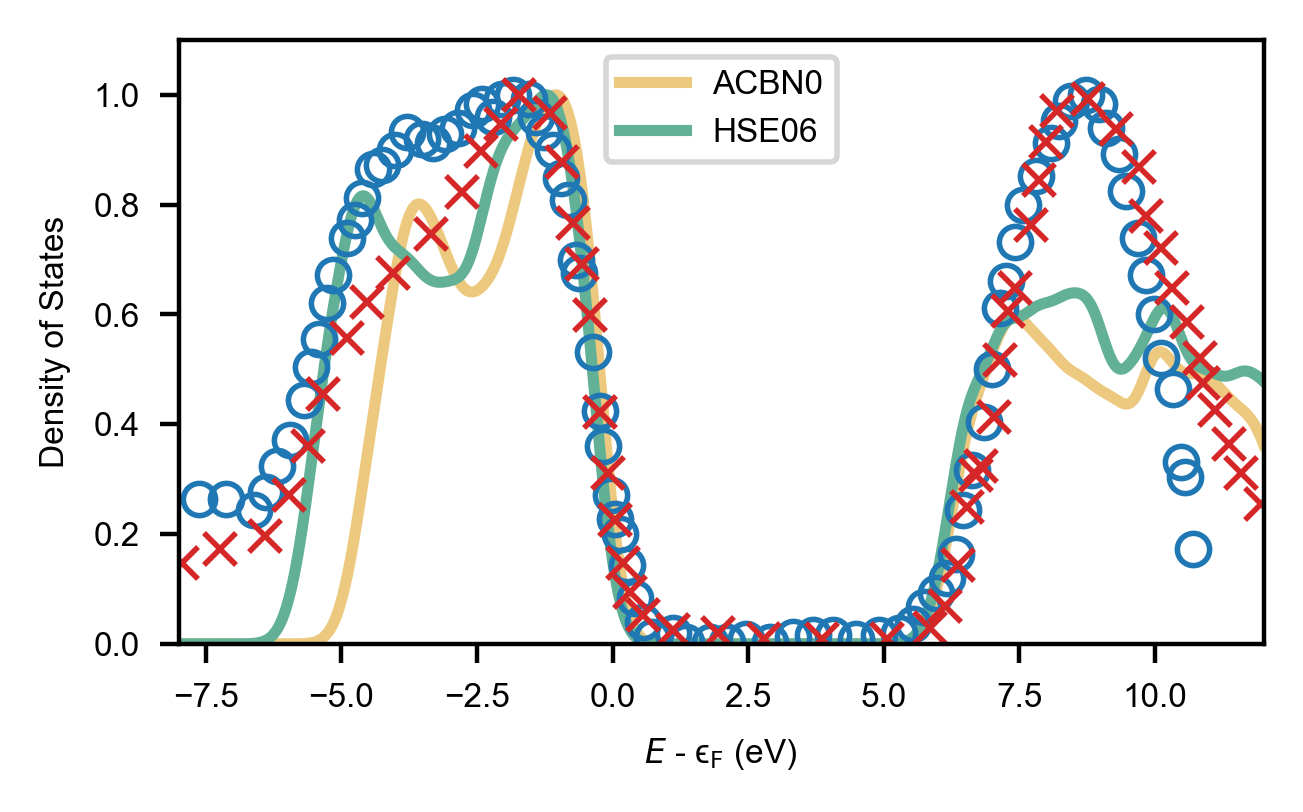}
    \caption{Comparison of DOS plots of the $M$ phase near the Fermi level for ACBN0 and HSE06 with experimental data. The red crosses represent measurements from ref.~\cite{Sayan04p7485} using spectroscopic ellipsometry, and the blue circles are from ref.~\cite{Bersch08p085114} measured by x-ray photoemission spectroscopy.}
    \label{expdos}
\end{figure*}

As ACBN0 and HSE06 predict similar band gaps, we analyze their DOS plots with the VBM aligned across a broad energy window as shown in Fig.~\ref{gwdos} and compare them with those obtained from the high-level $GW_0$ method~\cite{Jiang10p085119}.
All three methods yield comparable DOS profiles for states near the \note{band edges}.
Nevertheless, ACBN0 indicates that the Hf-4$f$ states are located between $-15$ and $-12$~eV, while HSE06 and $GW_0$ predict these states at lower energies, approximately 4.5~eV below. This difference is expected, as no Hubbard correction is applied to the Hf-$4f$ states in ACBN0. 
\notetwo{Interestingly, within the energy window from -6~eV to 0~eV, the DOS profile predicted by eACBN0 shows better agreement with $GW_0$ and HSE06 compared to ACBN0.}
We also compare the theoretical DOS results for the $M$ phase with available experimental data. As shown in Fig.~\ref{expdos}, both ACBN0 and HSE06 largely reproduce the main features of the experimental spectra. Based on the detailed comparisons above, ACBN0 emerges as a cost-effective method for describing the electronic structures of hafnia, particularly for states near the band edges.

\subsection{Self-consistent Hubbard $U$ parameters around vacancy}

As is inherent to the Hubbard model, the $U$ parameter is dependent on the local atomic environment and is specific to each Hubbard site. The self-consistently determined $U$ values in ACBN0 can serve as descriptors of local atomic environments, allowing us to probe the perturbations introduced by point defects such as oxygen vacancies.
To model a single oxygen vacancy in HfO$_2$, we optimized 2$\times$2$\times$2 supercells with one oxygen atom removed, followed by a single-point energy calculation using ACBN0 to determine the site-specific $U$ values. 

\begin{figure*}[h]
    \centering
    \includegraphics[]{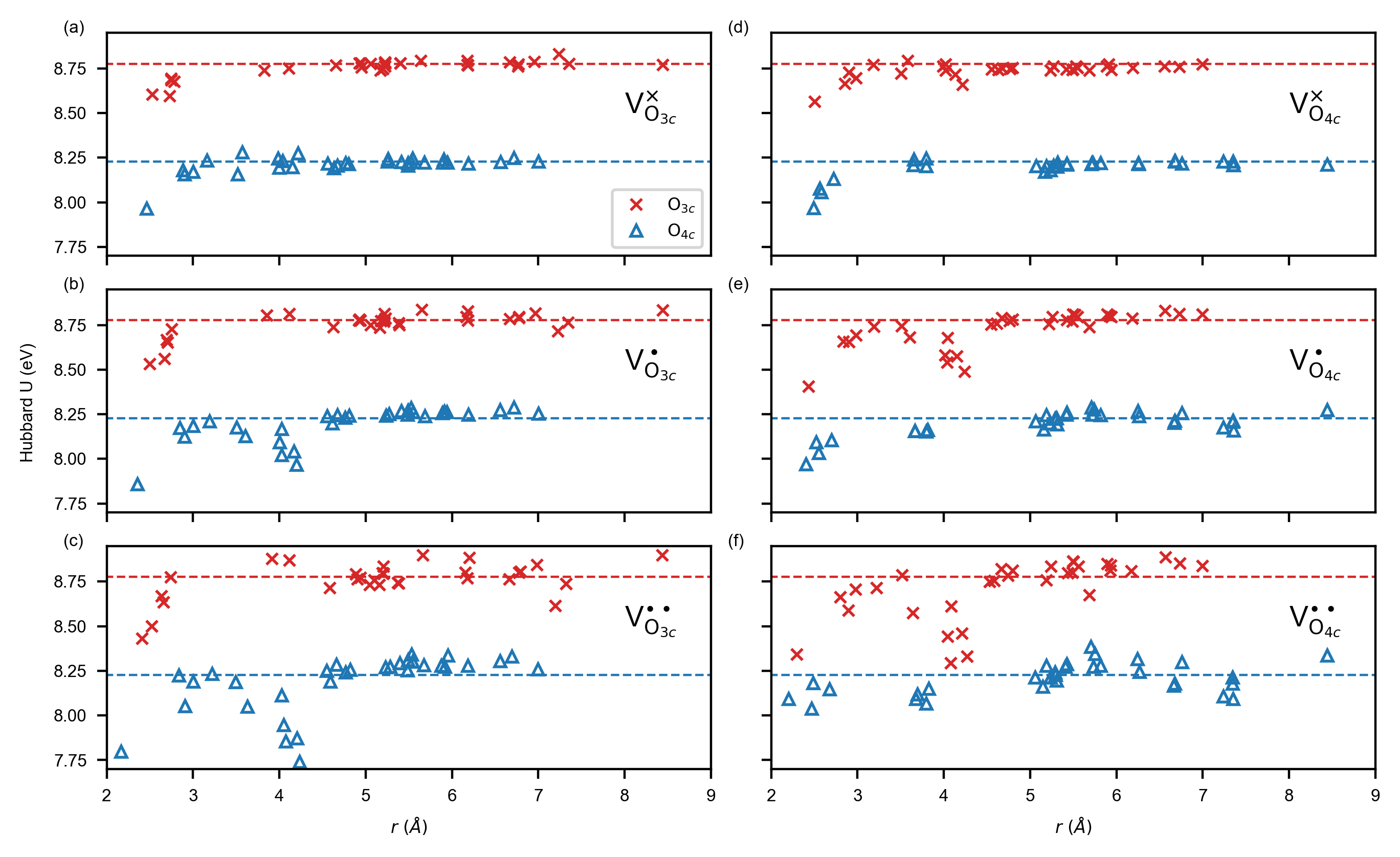}
    \caption{Variation of the magnitude of $U$ on O-2$p$ states predicted by ACBN0 with respect to the distance from 
    (a) charge-neutral V$_{\rm{O_{3c}}}^\times$, 
    (b) singly-charged V$_{\rm{O_{3c}}}^\bullet$, (c) doubly-charged V$_{\rm{O_{3c}}}^{\bullet\bullet}$, (d) V$_{\rm{O_{4c}}}^\times$, (e) V$_{\rm{O_{4c}}}^\bullet$, and (f) V$_{\rm{O_{4c}}}^{\bullet\bullet}$ in the $M$ phase. Red crosses represent the magnitude of $U$ on $2p$ orbits of O$_{\rm{3c}}$, and the blue triangles represent the magnitude of $U$ of $2p$ orbits of O$_{\rm{4c}}$. The dashed lines denote the $U$ values in pristine structures. }
    \label{uchange}
\end{figure*}

The oxygen-deficient $M$ phase features two types of oxygen vacancies, \Vop~and \Vonp, created by removing \Op~and \Onp, respectively. We plot the site-specific $U$ values of the O-$2p$ states as a function of the distance ($r$) of oxygen atom from the vacancy in different charge states (V$_{\rm O}^\times$, V$_{\rm O}^\bullet$, V$_{\rm O}^{\bullet\bullet}$) in Fig.~\ref{uchange}. 
For the case of V$_{\rm{O_{3c}}}$, we find that the $U$ values of \Op~atoms rapidly converge to their bulk values as $r$ increases, regardless of the vacancy's charge state.
In contrast, the $U$-$r$ relationships for \Onp~atoms exhibit more complex behavior. Specifically, for charged vacancies like V$^{\bullet}_{\rm{O_{3c}}}$ and V$^{\bullet\bullet}_{\rm{O_{3c}}}$, we observe a sharp decrease in the $U$ values for \Onp~atoms at $r \approx 4$ \AA, beyond which the $U$ values eventually return to the stoichiometric value. The situation is similar but subtly different for V$_{\rm{O_{4c}}}$. Here, the $U$ values of \Onp~atoms (rather than \Op~atoms) quickly recover their bulk values with increasing $r$. When the vacancy becomes charged, the $U$ values for \Op~atoms drop substantially at $r \approx 4$ \AA~from the vacancy. Similar trends are also observed in the PO and $T$ phases. 

\begin{figure*}[h]
    \centering
    \includegraphics[width=1.0\textwidth]{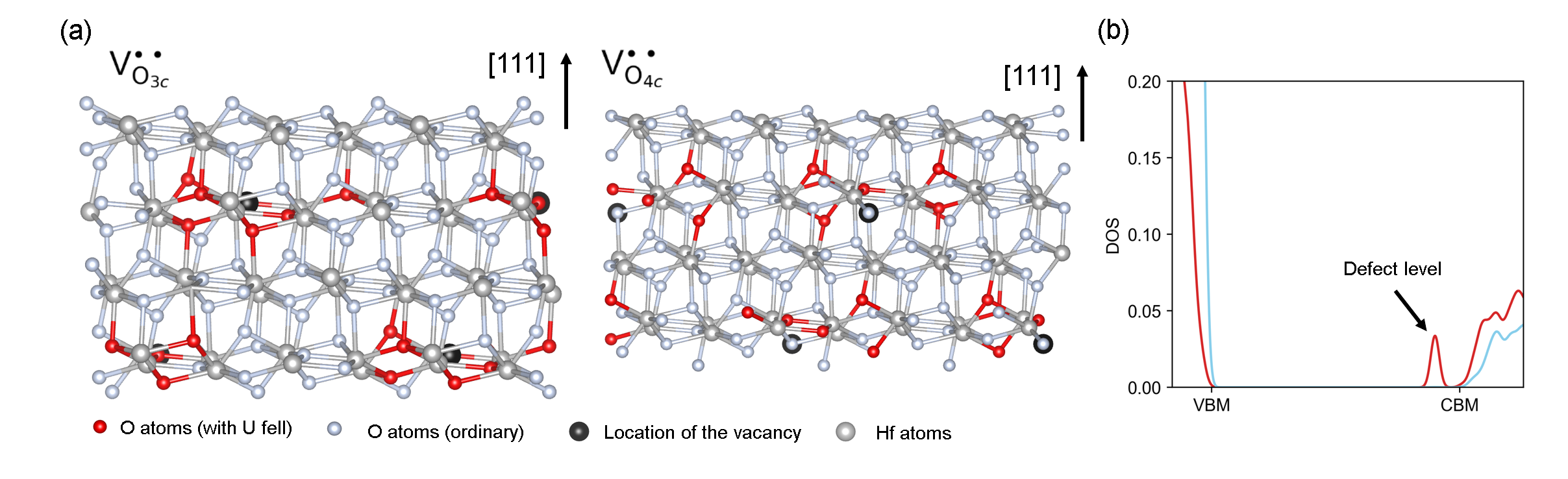}
    \caption{(a) Spatial distribution of oxygen atoms with significant variation in $U$ values, take the $M$ phase as an example. (b) Atom-resolved DOS. The red line corresponds to the oxygen atoms marked in red in (a), indicating a contribution at the defect level. In contrast, sky-blue colored oxygen atoms do not contribute to the defect level. }
    \label{ustru_pdos}
\end{figure*}

Our analysis reveals that oxygen atoms experiencing substantial variations in the $U$ value ($>$0.25 eV) are primarily distributed around the vacancy within the \{111\} plane, as illustrated in Fig.~\ref{ustru_pdos}(a). Although some oxygen atoms are closer to the vacancy, their $U$ values are less affected because they reside in neighboring \{111\} planes. This spatial distribution explains the notable drop in $U$ values at $r\approx 4$~\AA~for \Onp~near charged V$_{\rm{O_{3c}}}$ and \Op~near charged V$_{\rm{O_{4c}}}$. 
Furthermore,  DOS calculations indicate that these oxygen atoms exhibit strong hybridization with the defect level (see Fig.~\ref{ustru_pdos}(b)), contributing to their significant changes in $U$ values. These findings suggest the potential utility of site-specific $U$ values in elucidating the electronic structures of materials in real space. The observed spatial variations in $U$ values, particularly around defects, provide useful insights into local electronic environments.

\subsection{Defect levels in ferroelectric HfO$_2$}
\begin{figure*}[h]
    \centering
    \includegraphics[]{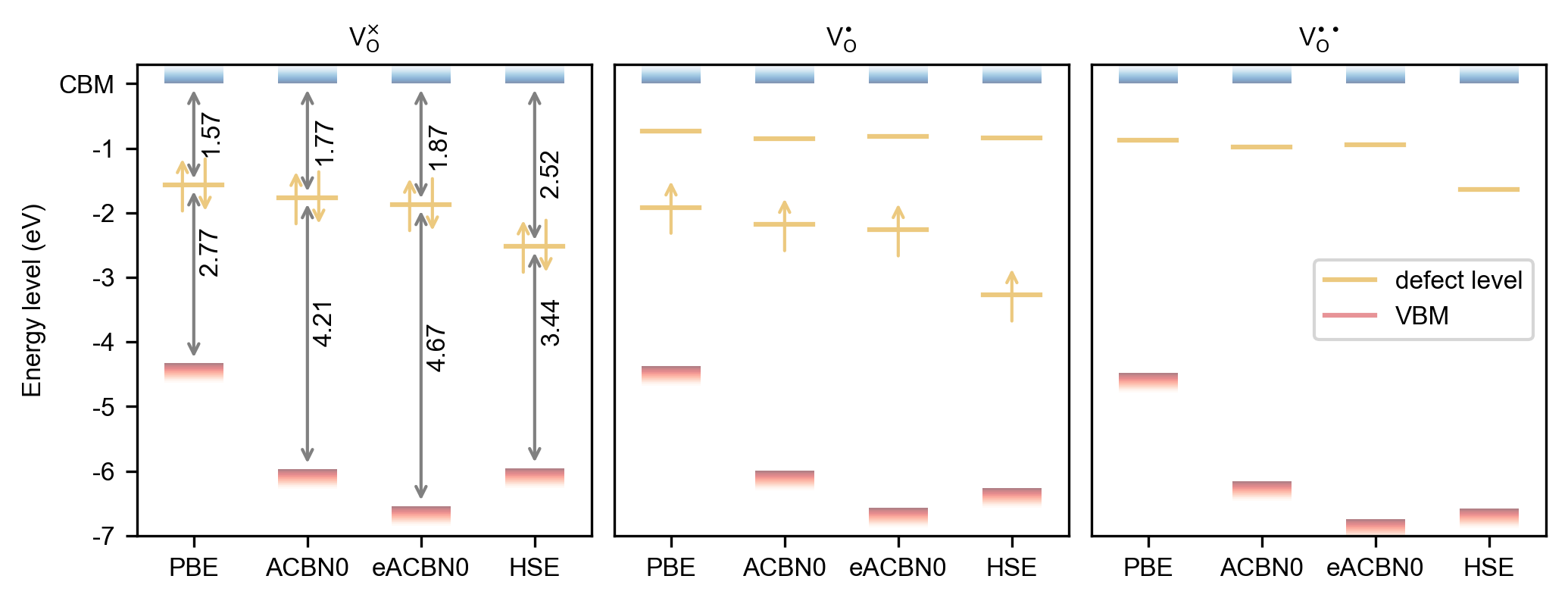}
    \caption{Relative energies of CBM, VBM, and defect levels calculated using PBE, ACBN0, eACBN0, and HSE06. For V$_{\rm{O}}^{\bullet}$, the spin-polarization is considered. The energy of the CBM is set to zero. Occupied states are illustrated by the yellow arrows.}
    \label{charge_state}
\end{figure*}

Figure \ref{charge_state} offers a comparative analysis of the relative energies of the VBM and the defect level in the PO phase of HfO$_2$ with V$_{\rm O_{\rm 3c}}$ of different charge states, treating the CBM as a reference energy level.
In the presence of V$_{\rm O}^\times$, the defect level is doubly occupied, corresponding to the electron localization at the vacancy site. The energy differences between the CBM and the defect level increase in the order of PBE, ACBN0, eACBN0, and HSE06, whereas eACBN0 predicts the largest difference between the VBM and the defect level.
The effect of the XC functional on the relative energy of the defect level remains largely consistent for singly-charged V$^{\bullet}_{\rm{O}}$ and doubly-charged V$^{\bullet\bullet}_{\rm{O}}$. Overall, ACBN0 and HSE06 yield similar relative positions for the defect level within the band gap.

\begin{figure*}[h]
    \centering
    \includegraphics[width=1.0\textwidth]{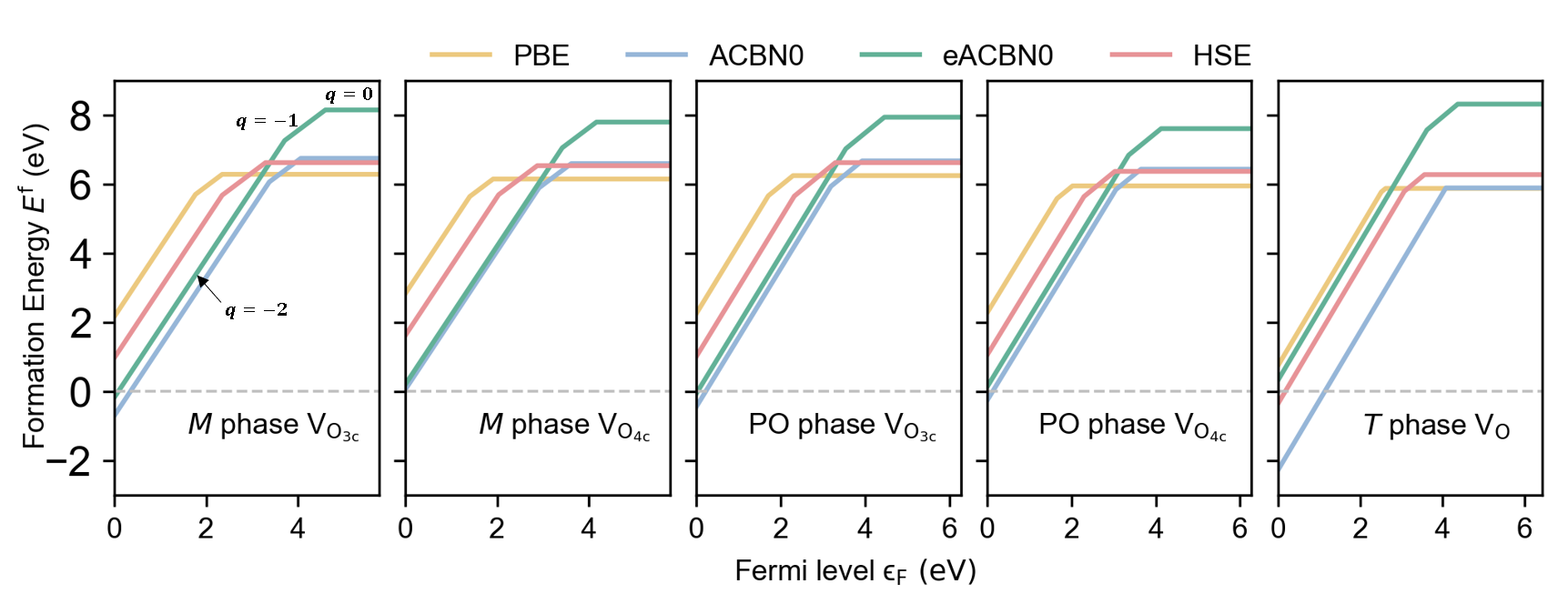}
    \caption{Oxygen vacancy formation energy as a function of Fermi level $\epsilon_{\rm{F}}$ for \Vop, \Vonp~in the $M$ phase, \Vop, \Vonp~in the PO phase, and for V$_{\rm{O}}$~in the $T$ phase.} 
    \label{FE}
\end{figure*}

We also investigate the formation energy of oxygen vacancies in hafnia polymorphs, with the results graphically presented in Fig.~\ref{FE}. 
Experimentally, the Fermi energy can vary between the VBM and the CBM, depending on conditions such as vacancy concentration and temperature. 
Therefore, it is common practice in defect formation energy calculations to treat the Fermi energy as a variable that spans from the VBM to the CBM.
Taking HSE06 predictions as a benchmark, our findings show that ACBN0 consistently delivers accurate results for neutral vacancies in $M$, PO, and $T$ phases, while the standard PBE functional predicts lower values. 
In contrast, eACBN0 tends to overestimate the formation energies by $\approx$1.5~eV.
This overestimation is primarily due to eACBN0 predicting a significant energy difference between the VBM and the defect level (see Fig.~\ref{charge_state}), resulting in vacancies with localized high-energy electrons, making them less stable.
When considering charged defects, the values from ACBN0 and eACBN0 are comparable, both lower than those computed with HSE06. We observe that all functionals indicate that the formation energy of V$^{\bullet\bullet}_{\rm{O_{3c}}}$ is lower than that of V$^{\bullet\bullet}_{\rm{O_{4c}}}$, aligning with previous research~\cite{Wei21p2104913}. It is noted that when the Fermi level is close to the VBM, ACBN0, and eACBN0 predict negative formation energies for charged vacancies, suggesting their possible spontaneous formations under suitable $p$-type doping conditions. 

\notetwo{Finally, we evaluated the computational efficiency of various functionals for single-point energy calculations of HfO$_2$ using Intel Xeon Platinum 8358 processors with 64 cores per node. For the 12-atom unit cell, 1 node was used. The PBE functional completed the calculation in 11.95 seconds, ACBN0 in 1 minute and 17.68 seconds, eACBN0 in 4 minutes and 55.86 seconds, and HSE06 in 12 minutes and 46.01 seconds. For a 2$\times$2$\times$2 supercell (96 atoms), 2 nodes were utilized. The PBE functional completed the calculation in 2 minutes and 21.26 seconds, ACBN0 in 18 minutes and 21.96 seconds, eACBN0 in 1 hour and 5 minutes, and HSE06 in 1 hour and 51 minutes. These results show that PBE is consistently the fastest. ACBN0, though slower, remains efficient and performs comparably to PBE for smaller systems. eACBN0, while slower than ACBN0, is significantly faster than HSE06, making it a practical alternative for improved efficiency.}

\section{Conclusions}
This study demonstrates that the DFT + $U$ method, utilizing Hubbard parameters computed with ACBN0, provides an efficient and accurate approach for predicting the physical properties of hafnia polymorphs. Our results show that ACBN0  closely reproduces the results obtained from HSE06 and $GW_0$ methods, and as well as experimental data, while being much less computationally demanding. Particularly, the analysis of the electronic states near the Fermi level indicates that both HSE06 and ACBN0 functionals yield predictions that align well with experimental data. 
In comparison, eACBN0 predicts a larger band gap and higher vacancy formation energy; however, the DOS profiles for states near the VBM exhibit closer alignment with the results from HSE06 and $GW_0$.
Overall, self-consistent Hubbard parameters can also serve as effective indicators of bond strengths and local changes in electronic structures in real space. 
Furthermore, our investigation of oxygen vacancy formation energies reveals that the ACBN0 functional, which incorporates environmentally dependent Hubbard parameters and local perturbations around defects, provides reliable predictions of relative energies. This study highlights the potential of DFT + $U$ with self-consistent Hubbard parameters for applications in transition metal oxides.  

\begin{acknowledgments}
We gratefully acknowledge Liyang Ma, Zhuang Qian, and Mohan Chen for the helpful conversations, and Changming Ke and Tianyuan Zhu for the manuscript preparation. This work is supported by National Natural Science Foundation of China (12074319) and Westlake Education Foundation. Y.-W.S. was supported by KIAS individual Grant (No. CG031509). W.Y. was supported by KIAS individual Grant (No. 6P090103). The computational resource is provided by Westlake HPC Center.
\end{acknowledgments}

\bibliography{SL}

\clearpage
\newpage

\begin{table*}[]
    \renewcommand{\arraystretch}{1.5}
        \caption{Lattice constants of hafnia polymorphs optimized with PBE and self-consistent Hubbard $U$ and $V$ parameters computed using ACBN0 and eACBN0. There are two different types of oxygen atoms, denoted as O$_{\rm{3c}}$ and O$_{\rm{4c}}$, in the $M$ and PO phases of HfO$_2$.}
    \begin{tabular}{c |c |c |c } 
        \hline {\ \ \ \ \ Phases\ \ \ \ \ } & {\ \ \ \ Lattice Constants (\rm{\AA})\ \ \ \ } & {\ \ \ \ \ \ \ ACBN0 $U$ (eV)\ \ \ \ \ \ \ }& {\ \ \ \ \ \ \ eACBN0 $U$ (eV)\ \ \ \ \ \ \ } \\
        \hline
        \multirow{3}{*}{
        $M$ phase
        } 
        & $a$ = 5.24 
        & $U$(Hf-5$d$) = 0.158
        & $U$(Hf-5$d$) = 0.210  \\
        & $b$ = 5.07 
        & $U$(O$_{\rm{4c}}$-2$p$) = 8.227  
        & $U$(O$_{\rm{4c}}$-2$p$) = 8.052  \\ 
        & $c$ = 5.17 
        & $U$(O$_{\rm{3c}}$-2$p$) = 8.776  
        & $U$(O$_{\rm{3c}}$-2$p$) = 8.510  \\ 
        \midrule  
        \multirow{3}{*}{
        PO phase
        } 
        & $a$ = 5.22  
        & $U$(Hf-5$d$) = 0.157  
        & $U$(Hf-5$d$) = 0.210  \\
        & $b$ = 5.00  
        & $U$(O$_{\rm{4c}}$-2$p$) = 8.248   
        & $U$(O$_{\rm{4c}}$-2$p$) = 8.036   \\ 
        & $c$ = 5.03  
        & $U$(O$_{\rm{3c}}$-2$p$) = 8.782 
        & $U$(O$_{\rm{3c}}$-2$p$) = 8.554 \\ 
        \midrule
        \multirow{3}{*}{
        $T$ phase
        }
        & $a$ = 5.16  
        & $U$(Hf-5$d$) = 0.181  
        & $U$(Hf-5$d$) = 0.235  \\
        & $b$ = 5.03  
        & $U$(O-2$p$) = 8.512   
        & $U$(O-2$p$) = 8.394   \\ 
        & $c$ = 5.03  
        & {-} & {-} \\ 
        
        \hline 
    \end{tabular}
    \label{Acbn0U}
\end{table*}

\clearpage
\newpage

\begin{table}
\renewcommand{\arraystretch}{1.5}
\caption{Decomposition of the energy difference into the one-electron, Hartree, exchange-correlation, Ewald, and Hubbard contributions.}
\begin{tabular}{l|ccc} 
\hline
\hline
\multicolumn{1}{c|}{$\Delta E$ = $E$(PO) $-$ $E$($M$) (eV)} & {\ \ PBE\ \ }  & {\ \ ACBN0\ \ } & {\ \ eACBN0\ \ } \\ 
\hline
one-electron $\Delta E$ & 23.96 & 23.89 & 23.91 \\
Hartree $\Delta E$ & $-$9.76 & $-$9.68 & $-$9.69 \\
XC $\Delta E$ & $-$0.03 & $-$0.03 & $-$0.04 \\
Ewald $\Delta E$ & $-$14.15 & $-$14.15 & $-$14.15 \\
\hline
Sum of above $\Delta E$ & 0.31 & 0.31 & 0.34 \\
\hline
Extended Hubbard $\Delta E$ & 0.00 & $-$0.07 & 0.08 \\
\hline

\hline
\hline
\multicolumn{1}{c|}{\ \ $\Delta E$ = $E$($T$) $-$ $E$($M$) (eV)\ \ \ } & {\ \ PBE\ \ }  & {\ \ ACBN0\ \ } & {\ \ eACBN0\ \ } \\
\hline
one-electron $\Delta E$ & 26.78 & 26.60 & 26.67 \\
Hartree $\Delta E$ & $-$10.59 & $-$10.40 & $-$10.47 \\
XC $\Delta E$ & $-$0.05 & $-$0.06 & $-$0.06 \\
Ewald $\Delta E$ & $-$16.09 & $-$16.09 & $-$16.09 \\
\hline
Sum of above $\Delta E$ & 0.63 & 0.62 & 0.64  \\
\hline
Extended Hubbard $\Delta E$ & 0.00 & 0.04 & 0.14  \\
\hline
\hline
\end{tabular}
\label{energy_separate} 
\end{table}

\clearpage
\newpage

\clearpage
\newpage

\clearpage
\newpage

\clearpage
\newpage

\clearpage
\newpage

\clearpage
\newpage

\clearpage
\newpage




\newpage

\clearpage
\newpage

\clearpage
\newpage

\end{document}